%% file: sample-manuscript.tex
\documentclass[acmtog]{acmart}
\AtBeginDocument{%
  \providecommand\BibTeX{{%
    \normalfont B\kern-0.5em{\scshape i\kern-0.25em b}\kern-0.8em\TeX}}}

\setcopyright{rightsretained}
\copyrightyear{2022}
\acmYear{2022}
\acmVolume{41}
\acmNumber{6}
\acmArticle{234}
\acmMonth{12} 
\acmDOI{10.1145/3550454.3555505}

\acmJournal{TOG}

\usepackage{multirow}
\usepackage{algorithm}
\usepackage{algpseudocode}
\usepackage{color,xcolor}
\usepackage{threeparttable}

\usepackage{soul}
\begin{document}

\title{ICARUS: A Specialized Architecture for Neural Radiance Fields Rendering}

\author{Chaolin Rao}
\email{raochl@shanghaitech.edu.cn}
\affiliation{
  \institution{ShanghaiTech University}
  \city{Shanghai}
  \country{China}
}
\affiliation{
\institution{GGU Technology Co., Ltd.}
\country{China}
}

\author{Huangjie Yu}
\email{yuhj@shanghaitech.edu.cn}
\affiliation{
  \institution{ShanghaiTech University}
  \city{Shanghai}
  \country{China}
}

\author{Haochuan Wan}
\email{wanhch@shanghaitech.edu.cn}
\affiliation{
  \institution{ShanghaiTech University}
  \city{Shanghai}
  \country{China}
}

\author{Jindong Zhou}
\email{zhoujd@shanghaitech.edu.cn}
\affiliation{
  \institution{ShanghaiTech University}
  \city{Shanghai}
  \country{China}
}

\author{Yueyang Zheng}
\email{zhengyy1@shanghaitech.edu.cn}
\affiliation{
  \institution{ShanghaiTech University}
  \city{Shanghai}
  \country{China}
}

\author{Minye Wu}
\email{minye.wu@kuleuven.be}
\affiliation{
  \institution{KU Leuven}
  \country{Belgium}
}

\author{Yu Ma}
\email{mayu@shanghaitech.edu.cn}
\affiliation{
  \institution{ShanghaiTech University}
  \city{Shanghai}
  \country{China}
}

\author{Anpei Chen}
\email{chenap@shanghaitech.edu.cn}
\affiliation{
  \institution{ShanghaiTech University}
  \city{Shanghai}
  \country{China}
}

\author{Binzhe Yuan}
\email{yuanbzh@shanghaitech.edu.cn}
\affiliation{
  \institution{ShanghaiTech University}
  \city{Shanghai}
  \country{China}
}

\author{Pingqiang Zhou}
\email{zhoupq@shanghaitech.edu.cn}
\affiliation{
  \institution{ShanghaiTech University}
  \city{Shanghai}
  \country{China}
}

\author{Xin Lou}
\email{louxin@shanghaitech.edu.cn}
\authornote{Corresponding author.}
\affiliation{
  \institution{ShanghaiTech University}
  \city{Shanghai}
  \country{China}
}
\affiliation{
\institution{GGU Technology Co., Ltd.}
\country{China}
}

\author{Jingyi Yu}
\email{yujingyi@shanghaitech.edu.cn}
\authornotemark[1]
\affiliation{
  \institution{ShanghaiTech University}
  \city{Shanghai}
  \country{China}
}

\renewcommand{\shortauthors}{Rao et al.}

\begin{abstract}

The practical deployment of Neural Radiance Fields (NeRF) in rendering applications faces several challenges, with the most critical one being low rendering speed on even high-end graphic processing units (GPUs). In this paper, we present ICARUS, a specialized accelerator architecture tailored for NeRF rendering. Unlike GPUs using general purpose computing and memory architectures for NeRF, ICARUS executes the complete NeRF pipeline using dedicated plenoptic cores (PLCore) consisting of a positional encoding unit (PEU), a multi-layer perceptron (MLP) engine, and a volume rendering unit (VRU). A PLCore takes in positions \& directions and renders the corresponding pixel colors without any intermediate data going off-chip for temporary storage and exchange, which can be time and power consuming. To implement the most expensive component of NeRF, i.e., the MLP, we transform the fully connected operations to approximated reconfigurable multiple constant multiplications (MCMs), where common subexpressions are shared across different multiplications to improve the computation efficiency. We build a prototype ICARUS using Synopsys HAPS-80 S104, a field programmable gate array (FPGA)-based prototyping system for large-scale integrated circuits and systems design. We evaluate the power-performance-area (PPA) of a PLCore using 40nm LP CMOS technology. Working at 400 MHz, a single PLCore occupies 16.5 $mm^2$ and consumes 282.8 mW, translating to 0.105 uJ/sample. The results are compared with those of GPU and tensor processing unit (TPU) implementations.

\end{abstract}

\ccsdesc[500]{Hardware~Integrated circuits}
\ccsdesc[300]{Hardware~Very large scale integration design}

\keywords{Neural radiance fields (NeRF), Neural rendering, hardware accelerator}

\begin{teaserfigure}
  \centering
  \includegraphics[width=\textwidth]{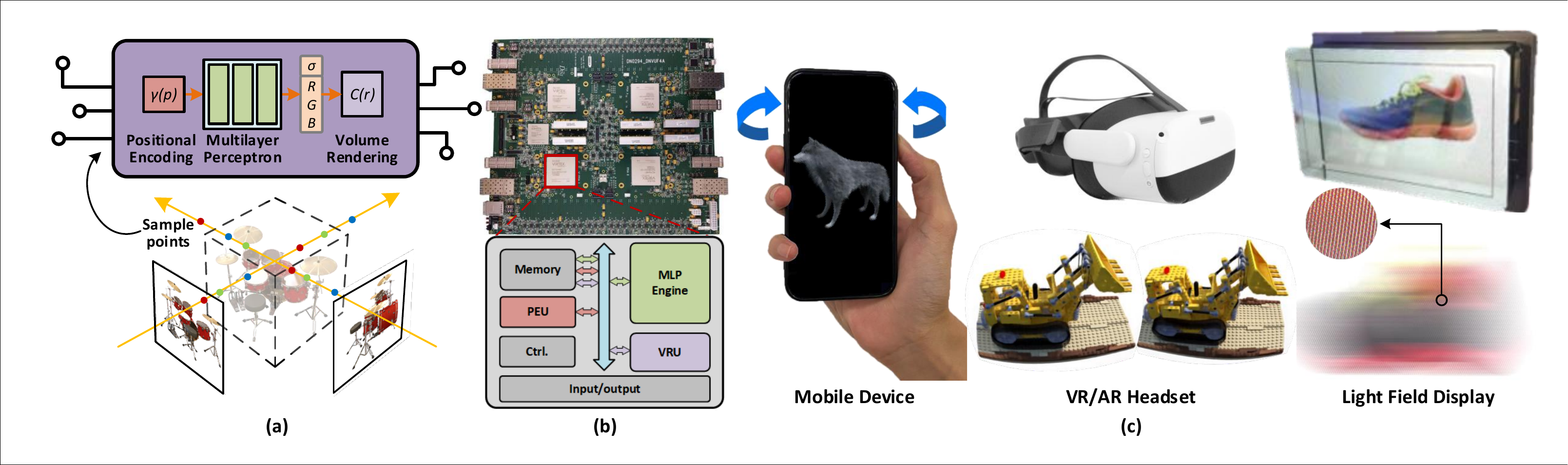}
  \caption{We demonstrate a specialized hardware architecture for NeRF-based rendering applications. (a) Our hardware design involves operations in NeRF-based rendering. (b) We validate our architecture using an FPGA platform. (c) Potential applications that can benefit from our architecture.}
  \Description{}
  \label{fig:Teaser2}
\end{teaserfigure}
\maketitle

\input{sections/Intro_new}

\input{sections/related-work}

\input{sections/design-philosophy}

\input{sections/system-design}

\input{sections/implementation}

\input{sections/discussion}

\bibliographystyle{ACM-Reference-Format}
\bibliography{sample-base}

\end{document}

%% file: sections/Intro_new.tex
\section{Introduction}
Advances on computer graphics algorithms are often trailed by their adoptions by hardware. Classic examples range from earlier z-buffer~\cite{catmull1974subdivision} for depth sorting~\cite{newell1972solution}, to geometry shaders~\cite{bailey2007glsl} for realizing various subdivision schemes~\cite{kazakov2007catmull, han2007tessellating, akenine2019real}, and to tailored microarchitectures (RDNA2 of AMD and Ampere of NVIDIA)~\cite{RDNA2, ampere} for accelerating ray tracing~\cite{meister2021survey, sanzharov2020survey}. With their focus anchored at producing physically correct rendering, latest hardware solutions still struggle to strike an intricate balance between real-time performance and high photorealism, especially for applications that can only afford lightweight graphics architecture. 

Recent neural rendering approaches provide an alternative, generation oriented solutions to traditional physical simulation. The seminal work of Neural Radiance Fields (NeRF)~\cite{nerf} implicitly reproduces the complete plenoptic function, i.e., the radiance along every ray, from sampled rays via tailored deep networks. The idea can be traced back to image-based modeling and rendering (IBMR) that first stores a set of captured photographs as a ray database and then employs tailored interpolation schemes to reconstruct the plenoptic function for view synthesis. Earlier IBMR techniques faced hurdles on both rendering quality and speed. For example, as occlusions between scene objects are inherently discontinuous and hence difficult to interpolate, they can cause severe aliasing artifacts unless ultra-dense ray samples are captured or highly accurate geometry is accessible, both prohibitively difficult under practical settings. In addition, storing ray samples as textures leads to excessive texture memory access at the rendering stage, limiting the performance of modern GPUs. 

To mitigates the occlusion problem, NeRF employs a volume rendering model that first estimates density (occupancy) changes along each ray and then aggregates the color accordingly. It uses a multi-layer perceptron (MLP) neural net to reconstruct, store, and query the complete radiance field. In fact, its MLP-based representation serves as a de facto compressor and interpolant: the standard NeRF uses around 1,200,000 parameters of a total size ~4.6MB to provide a continuous representation of the radiance field. A key bottleneck of NeRF is training the network. Various acceleration schemes have emerged based on tailored data structures \cite{liu2020neural,yu2021plenoxels, yu2021plenoctrees}, smart training strategies \cite{kangle2021dsnerf,sun2021direct}, and hardware features on graphics architecture \cite{muller2022instant}, reducing training time from hours to seconds. Reduction of training time comes at the cost of high space usage where advanced and expensive graphics hardware is required to store the intermediate data to enable real-time rendering. 

In this paper, we present ICARUS, a specialized architecture tailored for MLP-oriented neural rendering such as NeRF. GPUs use general purpose architectures for the computation of MLP and other neural rendering steps, along with additional components and off-chip memory for storing and exchanging intermediate data. ICARUS, in contrast, uses dedicated plenoptic cores (PLCore) as key computation components to execute the complete volume rendering pipeline without off-chip data exchanges and temporary storage. Each PLCore consists of a positional encoding unit (PEU), an MLP engine, and a volume rendering unit (VRU). It takes in positions \& directions and outputs the corresponding pixel colors with all computation and storage on-chip. This allows ICARUS to implement NeRF rendering in high parallelism with little overhead of control and off-chip memory access, significantly reducing time and power consumption. In particular, for MLP computation, we transform the fully connected operations to approximated reconfigurable multiple constant multiplications (RMCMs). This further leads to about 1/3 hardware complexity reduction compared to conventional multiply–accumulate (MAC)-based approaches.

We demonstrate a prototype single-core version ICARUS using Synopsys HAPS-80 S104, a field programmable gate array (FPGA)-based prototyping system for large-scale integrated circuits and systems. We evaluate the power-performance-area (PPA) of a PLCore using 40nm LP CMOS technology. Working at 400 MHz, a PLCore occupies 16.5 $mm^2$ and consumes 282.8 mW, translating to 0.105 uJ/sample. 
Apart from the original NeRF, we also evaluate implicit signed distance function (SDF) on ICARUS, to extract 3D geometry from the trained MLP via isosurface polygonisation for subsequent geometry processing tasks, useful in both 3D object scanning and other applications such as computerized tomography (CT) and magnetic resonance imaging (MRI) visualizations. With known 3D geometry, we also demonstrate bypassing the radiance aggregation component in ICARUS to directly conduct surface light field (SLF) rendering. We also develop tailored instructions for ICARUS to support NeRF rendering on a variety of displays. As a lightweight architecture, ICARUS aims to enable the displays to "fly" without tethering them to a mainframe and at the same time reduce the power consumption to avoid being "burned", critical for future immersive experiences.

%% file: sections/related-work.tex
\newcommand{\boldstart}[1]{\noindent\textbf{#1}}
\newcommand{\boldstartspace}[1]{\vspace{0.1in}\noindent\textbf{#1}}

\section{Related Work}

\begin{figure*}
    \centering
    \includegraphics[width=0.95\textwidth]{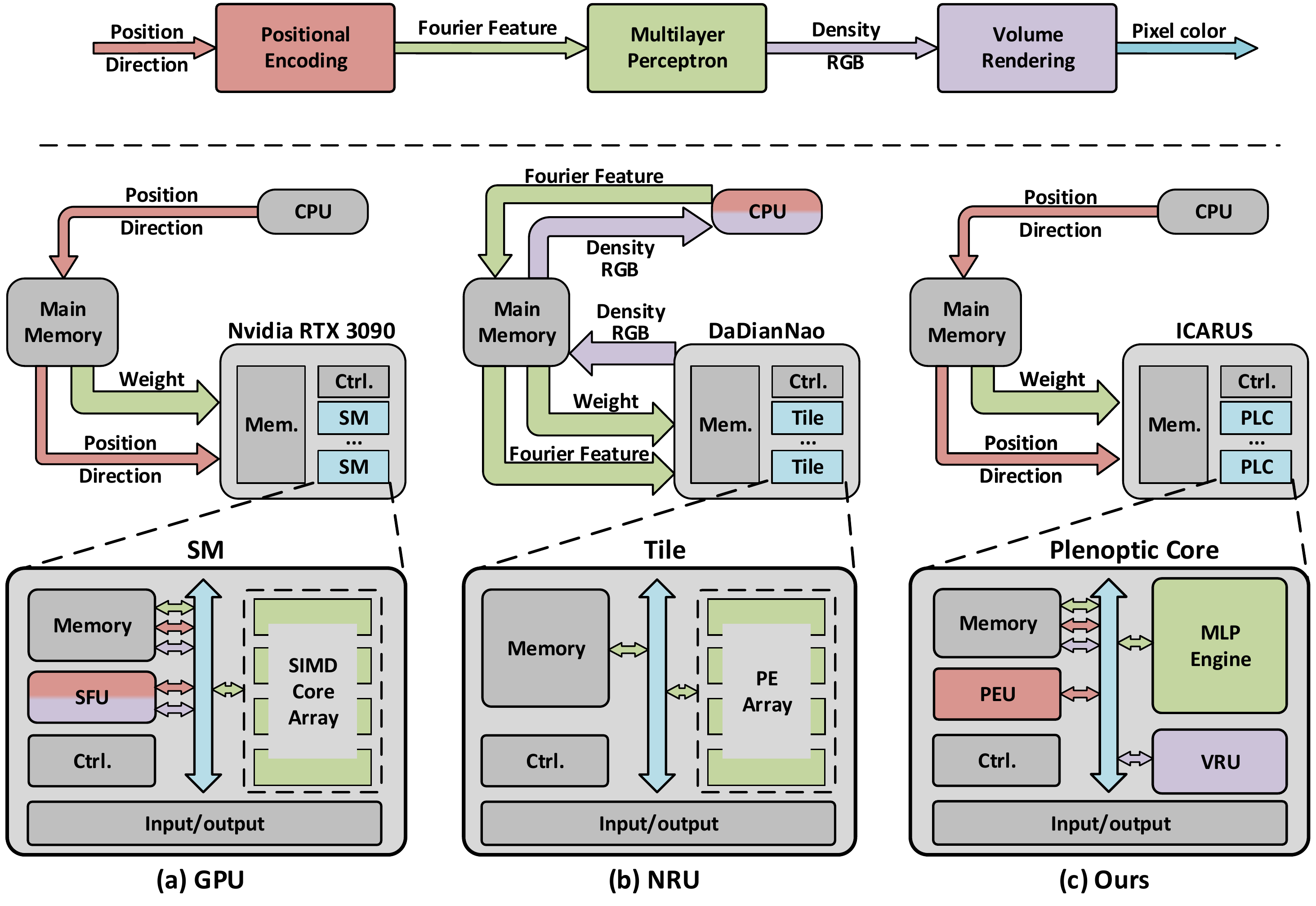}
    \caption{Illustration of mapping NeRF to (a) GPU (b) NPU and (c) ICARUS. For GPU-based NeRF implementation, positional encoding and volume rendering are executed in the SFU, and MLP is mapped to SIMD core arrays. Intermediate data is exchanged using off-chip memory. For NPU-based NeRF implementation, positional encoding and volume rendering are execuated in CPU and MLP is mapped to PE arrays. Intermediate data is also exchanged using off-chip memory. ICARUS executes the complete NeRF pipeline using dedicated plenoptic cores (PLCore) consisting of a PEU, an MLP engine, and a VRU.}
    \label{fig:computing_options}
\end{figure*}

ICARUS, as an academic prototype, aims to provide a dedicated architecture for neural radiance fields rendering. In many ways, it follows the footstep of the ray tracing hardware that started with a well studied algorithm and eventually landed to the hardware, with strong support from both academia and industry. As a neural rendering architecture, ICARUS also borrows design strategies from deep neural network chips to balance between power, memory, and bandwidth. In this section, we discuss the most relevant works in respective fields.

\boldstartspace{Ray Tracing Hardware.} Dedicated graphics hardware has unexceptionally come from its academic prototypes. Earlier volume rendering accelerator had been under heavy research in academia before powerful GPUs became accessible from industry. The VolumePro system\cite{VolumePro} focused on accelerating ray casting whereas its followup efforts were largely devoted to customizing ray-tracing hardware. Among them, one of best recognized academic breakthrough is SaarCOR (Saarbrücken's Coherence Optimized Ray Tracer) \cite{SaarCOR}, an FPGA-based prototype ray-tracing accelerator, which was later extended to the programmable Ray Processing Unit (RPU) \cite{RPU} and the Dynamic Ray Processing Unit (DRPU) \cite{DRPU} on products. Academic results have further inspired various microarchitectures based solutions by the GPU industry to achieve near real-time performance \cite{TRaX,HART} as well as to maintain energy-efficiency on mobile devices \cite{RayCore,MRTP,TandI,MRTP1}. They have all eventually landed on today's high-end commercial GPUs in the form of dedicated ray tracing cores, from NVIDIA's RTX series to AMD's RDNA architecture, and to Imagination Technologies' mobile PowerVR Photon. The development of ICARUS in this paper hopes to continue this legacy of technology transformation, stemming from academic prototypes and landing on industrial product, but this time in neural rendering.

\boldstartspace{Neural Network Accelerators.}
Neural networks, especially convolutional neural networks (CNNs), are widely used in many machine leaning applications. The seminal Tensor Processing Unit (TPU) \cite{TPU} from Google adopts a systolic array based design. While its first generation had merely focused on inference tasks (so is ICARUS), later TPU generations, i.e., TPUv2, support both training and inference, showing numerous successes in science and engineering. In fact, TPU is general enough to directly support neural rendering tasks such as NeRF \cite{jaxnerf2020github}. However, even with multiple TPUs, brute-force implementation cannot yet reach real-time performance (0.35 seconds per frame, with 128 TPUv2). Our goal is to develop a tailored rather than general architecture like the TPU. 

Light weight neural network solutions have also been extensively studied in the past decade, e.g., DianNao ~\cite{DianNao} and its subsequent extensions ~\cite{DaDianNao,ShiDianNao,PuDianNao}, Eyeriss ~\cite{Chen2017Eyeriss} and the follow-up work Eyeriss v2 ~\cite{Eyerissv2}, and the Thinker ~\cite{Thinker,ThinkerII,ThinkerS} series, etc.
Existing CNN accelerators have mainly focused on data flow optimization to eliminate, at least partially, the bottleneck caused by data movement ~\cite{DianNao,Chen2017Eyeriss}. They also emphasize more on the optimization of the convolution operation where a kernel is often convolved with different parts of input feature maps and a specific part of a feature map is convolved with different kernels. Therefore, the architecture design has mainly focused on weight sharing and/or activation strategies ~\cite{Moons,Thinker,DSIP} for improving the performance and at the same time reducing power consumption. For neural radiance fields rendering, whether in its original NeRF version or in consecutive optimized accelerated versions~\cite{mueller2022instant}, fully-connected MLPs instead of CNNs serve as the backbone. Since MLPs are fully connected, a weight is multiplied with only one input neuron and a neuron is multiplied with only a set of weights.
Therefore, directly mapping MLP-based rendering scheme to existing CNN accelerators causes efficiency degradation since the computation patterns are different. Fig. \ref{fig:computing_options}(b) illustrates the mapping of NeRF's MLP to a CNN accelerator architecture (DaDianNao~\cite{DaDianNao} in this case).

\boldstartspace{Algorithm-level Optimizations.}
Our ICARUS prototype follows largely the pure MLP design as in the original NeRF. The earliest version of NeRF was implemented on NVIDIA's GPU. Fig. \ref{fig:computing_options}(a) illustrates running NeRF on GPUs. However, as a general purpose computation platform with high power consumption (350 W for NVIDIA RTX 3090), GPU still cannot achieve real-time performance.

A number of algorithm-level techniques have been proposed to accelerate rendering and in some cases training of NeRF ~\cite{ yu2021plenoctrees, garbin2021fastnerf, yu2021plenoxels, mueller2022instant}. These approaches adopt a hybrid representation where the space is partitioned into voxels with well-designed data structures to reduce MLP layers (e.g., down to 2) to accelerate sampling fetching. \cite{luo2021convolutional} leverages coarse explicit scene geometry priors to guide point sampling along with rays and hence reduce network inference computation whereas \cite{liu2020neural} uses Octree in its hybrid representation to avoid redundant MLP queries in free space. \cite{garbin2021fastnerf} compactly caches a deep radiance map from trained MLPs in voxels, preventing network inference. Similarly, \cite{yu2021plenoxels} and~\cite{yu2021plenoctrees} realize view-dependent effects by adopting spherical harmonics and store coefficients on a sparse voxel grid or an Octree structure respectively. KiloNeRF~\cite{reiser2021kilonerf} uses thousands of tiny MLPs to represent parts of the scene so as to reduce the cost of network queries for sample points. Although effective, these techniques unanimously rely on the GPU features for trading space for time and therefore require very large storage (1.9GB for PlenOctrees). 
Perhaps the most exciting advance is Instant-NGP ~\cite{mueller2022instant} which adopts a multi-resolution encoding scheme coupled with voxel hashing and manages to simplify the original deep MLP to a compact one, reducing training time from hours to seconds and achieving real-time rendering. Nonetheless, its reliance on GPU hardware features is more extensive. Our pure MLP-based ICARUS provides an alternative solution to achieve high performance rendering without the GPU reliance where concurrent and future extensions may be implemented with proper extensions.

%% file: sections/design-philosophy.tex
\section{Design Philosophy}

\begin{figure*}[t]
    \centering
    \includegraphics[width=\textwidth]{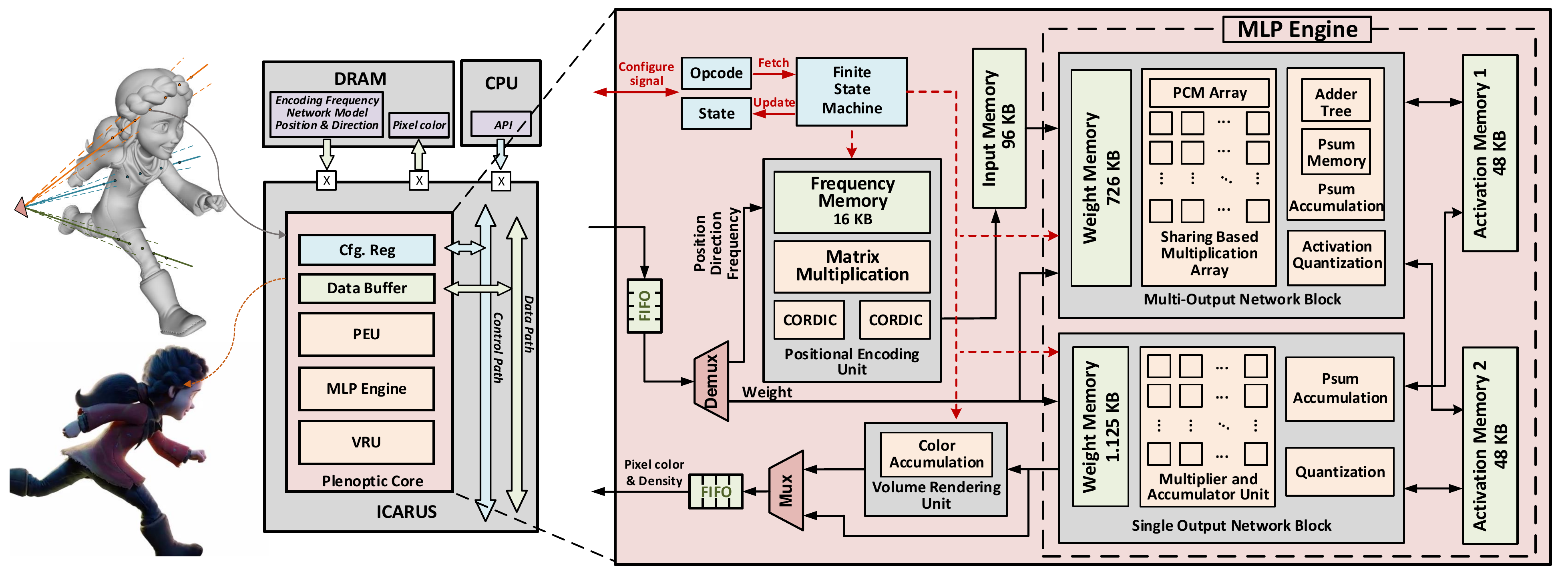}
    \caption{Overall architecture of the proposed ICARUS. The main computation components in ICARUS is PLCore. For NeRF rendering, a batch of sample points are processed by the PLCore, where the whole NeRF pipeline for a ray is completed inside the PLcore, i.e., a PLCore takes in positions \& directions and renders the corresponding pixel colors without any intermediate data going off-chip for temporary storage and exchange.}
    \label{fig:overallSys}
\end{figure*}

\subsection{Computation Analysis}
This section analyzes the computation in NeRF-based rendering tasks. 
NeRF implicitly models a continuous plenoptic function $F_{NeRF}:(\boldsymbol{p,d})\rightarrow(\boldsymbol{c},\sigma)$ with an MLP, which is trained using a set of images with known poses. By querying the MLP at a position $\boldsymbol{p} \in \mathbb{R}^3$ from a specific unit-norm viewing direction $\boldsymbol{d}\in \mathbb{R}^2$, the corresponding density $\boldsymbol{\sigma}$ and view-dependent color $\boldsymbol{c}$ at that position can be recovered. To render a pixel, a number of 3D positions, denoted by $(\boldsymbol{p}_1,\boldsymbol{p}_2,...,\boldsymbol{p}_N)$, are sampled along the ray that passes through the pixel. The plenoptic function $F_{NeRF}$ has to be evaluated for each sample point $\boldsymbol{p}_i$ to generate the corresponding color $\boldsymbol{c}_i$ and density $\sigma_i$. Direct volume rendering \cite{volume_rendering} is then applied to generate the color of the pixel.

As we can calculate, to render an image with millions of pixels, the plenoptic function $F_{NeRF}$ is evaluated $N$ times for each pixel, translating to tens or even hundreds of millions of MLP queries depending on the number of $N$. For illustration, the original NeRF samples 192 points along each ray. To render an $800\times800$ image, the MLP has to be queried $800\times800\times192=122880000$ times. This is the reason why it takes seconds to render a single frame even on modern high-end GPUs. 

The evaluations of MLP for different sample points are independent of one another. This means that they can be paralleled to improve the rendering speed, given sufficient computing resources. Parallel processing can be performed on the sample point-level or pixel (ray)-level. In addition, though not very critical in terms of computation, there are another two important steps in the NeRF-based pipeline: positional encoding which transforms positions \& directions into a higher dimensional feature space and volume rendering which integrates the colors of sample points to render the final pixel color. Due to the dimension increment, the amount of data for positions and directions is expended by 20 times and 8 times (for the original NeRF), respectively, after positional encoding. For example, to render an $800 \times 800$ image with 192 sample points, we only need to load in $800\times 800 \times 192 \times 6 \times 2B \approx 1.37$ GB if we include positional encoding in our system other than 19.22 GB if we do not (2 Bytes for each input data in our calculation). On the contrary, volume rendering compresses the amount of data by integrating the colors and densities of all sample points to generate the RGB values of the rendering pixels. For the fine rendering step with 128 sample points, the amount of data for an image from the MLP is about $800 \times 800 \times 128 \times 4 \times 2B \approx 625$ MB, which reduces to 3.66 MB after volume rendering.

\subsection{Design Decisions}

In this section, we present the design decisions for ICARUS. The first and most important decision is to finish the entire NeRF pipeline in a single PLCore without any intermediate data going off-chip for temporary storage and exchange, so as to save computation time and power consumption. To achieve this, we need to design a dedicated architecture which takes in positions \& directions and outputs final pixel colors. Moreover, though NeRF-based rendering algorithms share the same general flow consisting of positional encoding, MLP and volume rendering, they may differ in specific implementation of each module. Therefore, the dedicated architecture needs to have the flexibility to support different implementations of these modules. 

Another decision is to optimize the implementation of MLP for computational complexity reduction. Therefore, we use fixed-point instead of floating-point in ICARUS. It is well-known that floating-point computations are generally more complex than their fixed-point counterparts. Since NeRF is computationally demanding, it is very hard, if not impossible, to meet the performance and energy-efficiency requirement for mobile applications using floating-point computation. Apart from fixed-point computation, the fault-tolerant property of neural networks can be utilized to further reduce the computational complexity of NeRF.

%% file: sections/system-design.tex
\section{System Design}

\subsection{Overall System Architecture}

Fig. \ref{fig:overallSys} illustrates the overall architecture of ICARUS. For a rendering system consisting of CPU, dynamic random access memory (DRAM) and ICARUS, the inputs, i.e., network model, encoding frequency and positions \& directions are stored in DRAM and fed to ICARUS under the control of CPU during runtime. An on-chip network is used to dispatch input data to the destination PLCore. After loading the network model, positions \& directions are continuously streamed into ICARUS for NeRF processing and final rendered pixel colors are streamed out through the data path. Inside ICARUS, the sample points of a cluster of rays, comprising a sample point batch, are processed by the same PLcore, where the whole NeRF pipeline is completed inside the PLcore without any intermediate data going off-chip for temporary storage and exchange.

The detailed structure of a PLCore is illustrated in the right half of Fig. \ref{fig:overallSys}. According to the NeRF pipeline, we design a PLCore consisting of a PEU, an MLP engine and a VRU, where specific rendering steps are executed by the corresponding dedicated hardware module. The PEU transforms the input positions \& directions into higher dimensional feature space based on the input encoding frequencies. The encoded positions \& directions are fed to the MLP engine to render the colors $\boldsymbol{c}=(r,g,b)$ and densities $\sigma$ of the sample points, which are further consumed by the VRU to generate the final pixel colors. On-chip static random access memory (SRAM) blocks are used for temporary storage of weights, input positions \& directions and intermediate activations during the runtime. The execution of the NeRF pipeline on a PLCore is controlled by the internal finite state machine based on the input processing instructions.

\subsection{Positional Encoding Unit}

\begin{figure}[t]
    \centering
    \includegraphics[width=0.47\textwidth]{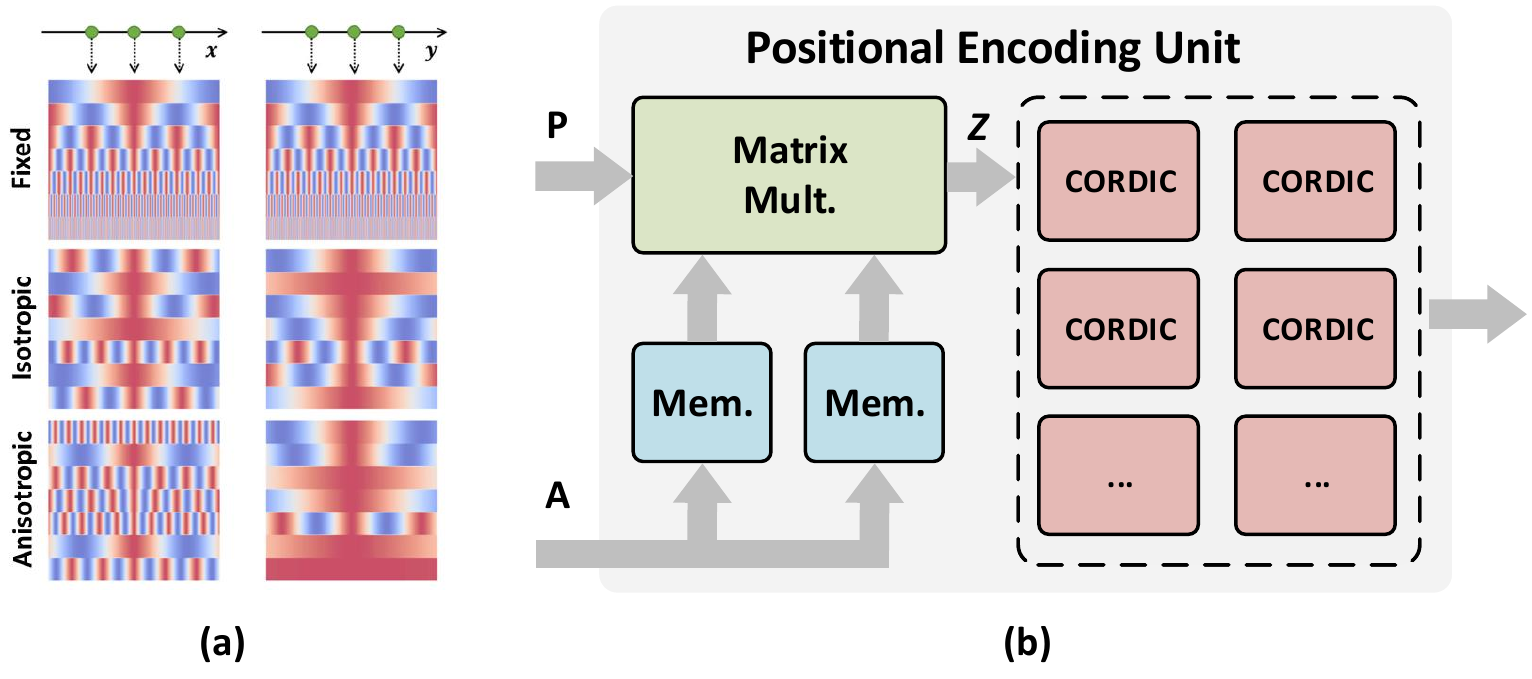}
    \caption{(a) Three different types of frequency patterns.   (b) Overall structure of the PEU. ($A$: Frequency matrix. $P$: Input positional data matrix.  $Z$: Results of the matrix multiplication of $AP$   )   }
    \label{fig:PEpic}
\end{figure}

Previous researches have revealed that the feed-forward neural networks tend to learn low frequency signals~\cite{nerf,zhong2019reconstructing,SpectralBias}, leading to over-blurred rendered images. One technique to overcome such difficulty is to map raw input coordinates into a relatively higher dimensional feature space through Fourier features mapping as
\begin{equation}
    \phi(x; A) = [\cos A^T x, \sin A^T x]
\end{equation}
This procedure is called positional encoding in NeRF-based neural rendering. Afterward, a neural network takes in transformed features instead of raw coordinate vectors.

It is easily perceived that the selection of the frequency matrix $A$ is critical for tuning the smoothness of network outputs. Shown in Fig.~\ref{fig:PEpic}(a) are different features (frequency matrices) used in different neural rendering algorithms, e.g., fixed frequency for NeRF rendering, isotropic random Fourier features for encoding implicit geometries \cite{ffm}, and anisotropic random Fourier features for neural image-based rendering of implicit geometries \cite{affm}.

In ICARUS, we develop a universal PEU for computing different kinds of feature mapping functions. Fig.~\ref{fig:PEpic}(b) illustrates the architecture of the PEU in a PLCore. The frequency matrix $A$, which is the same for all sample points in a particular rendering task, is stored in local memories. During the runtime, positions \& directions are streamed into the matrix multiplication unit, in which they are multiplied with the frequency matrix $A$ in a pipelined manner. The resultant intermediate vector $z$ is further fed to the Coordinate Rotation Digital Computer (CORDIC) array to generate the final transformed features, i.e., $sin(z) $ and $cos(z)$.

We design PEU to meet the requirement of inner product operations with two different sizes, i.e., ($\mathbb{R}^{3}$ and $\mathbb{R}^{6}$), for different positional encoding algorithms. For the case of $\mathbb{R}^{3}$, both frequency matrix data and position \& direction data are of 3-dimensional, meaning that one block memory with size $3\times128$ is enough to store the frequency matrix. For the inner product computation, a 3-stage cascaded MAC unit is used to perform three multiplications and two additions in serial. 
For the case of $\mathbb{R}^{6}$, instead of doubling the size of memory to $6\times128$ to maintain the same performance, we use two separate $3\times128$ memory blocks. Since two memory blocks can be accessed independently, the normal $\mathbb{R}^{3}$ frequency matrix or the first half of the $\mathbb{R}^{6}$ frequency matrix is loaded into the first memory block, and the other memory block stores only the second half of the $\mathbb{R}^{6}$ frequency matrix. In $\mathbb{R}^{6}$ computation mode, two memory blocks work simultaneously. While in $\mathbb{R}^{3}$ computation mode, the second memory is set to a sleep mode for power reduction. Moreover, the inner product results go through all 6 stages of the cascade MAC during $\mathbb{R}^{6}$ computation, while in $\mathbb{R}^{3}$ calculation, the later 3 stages are bypassed.

\begin{figure}[t]
    \centering
    \includegraphics[width=0.45\textwidth]{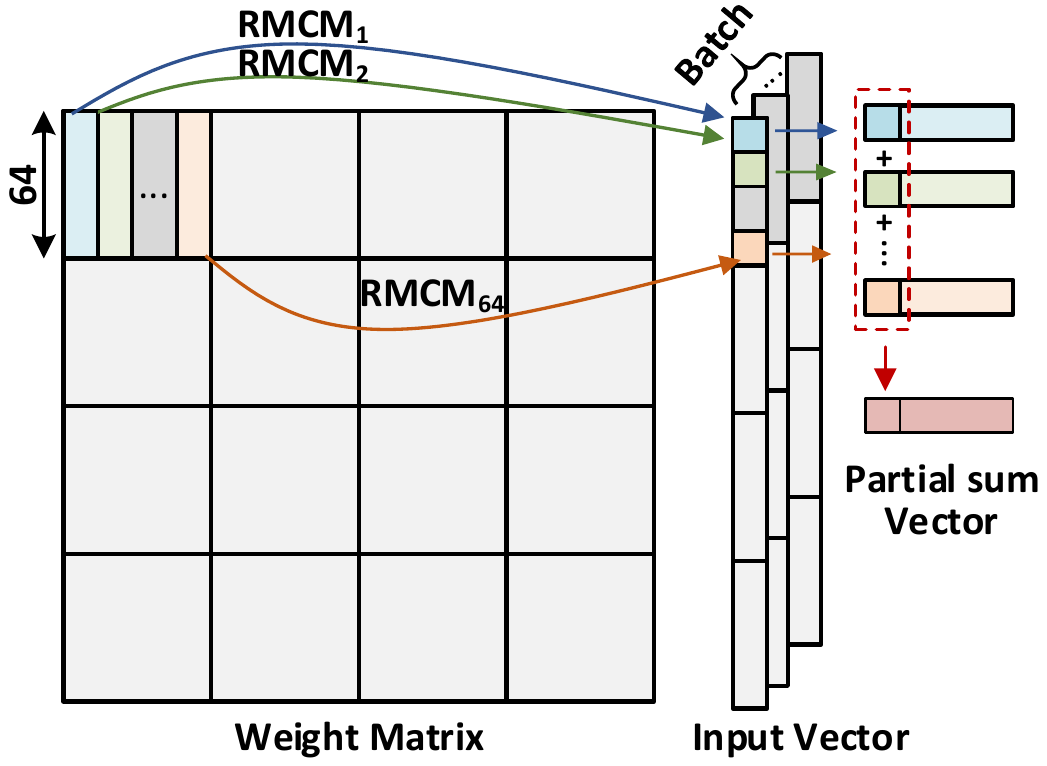}
    \caption{Overall computation flow in the MLP engine.}
    \Description{!}
    \label{fig:computation_flow}
\end{figure}

\subsection{MLP Engine}
In the NeRF pipeline, the MLP is the most computation intensive component because tens of millions of MLP query leads to tremendous amount of MAC operations. Therefore, we design a dedicated MLP engine to accelerate the computation of MLP. A typical MLP layer can be expressed as 
\begin{equation}
    \boldsymbol{y} = f({W}\boldsymbol{x}+\boldsymbol{b})
\end{equation}
where $\boldsymbol{x}$, ${W}$, and $\boldsymbol{b}$ are the input neuron vector, weight matrix and bias vector, respectively, and $f$ denotes the activation function. As an example, ${W}$ is a $256\times 256$ matrix in the original NeRF. 

The overall architecture of the MLP engine is illustrated in Fig.\ref{fig:overallSys}, which consists of a multi-output network block (MONB), a single output network block (SONB) and two activation memory blocks. The MONB block is responsible for the computation of hidden layers in an MLP where multiple output neurons are generated, while the SONB block is designed to compute the output layer of MLPs in NeRF, which generates the density or color of sample points. We design a dedicated SONB block because when mapping the output layer to MONB, most of the computation resources are wasted. We use a MAC array with 64 general multipliers to implement the fully-connected operations in SONB.

\begin{figure}[t]
    \centering
    \includegraphics[width=0.47\textwidth]{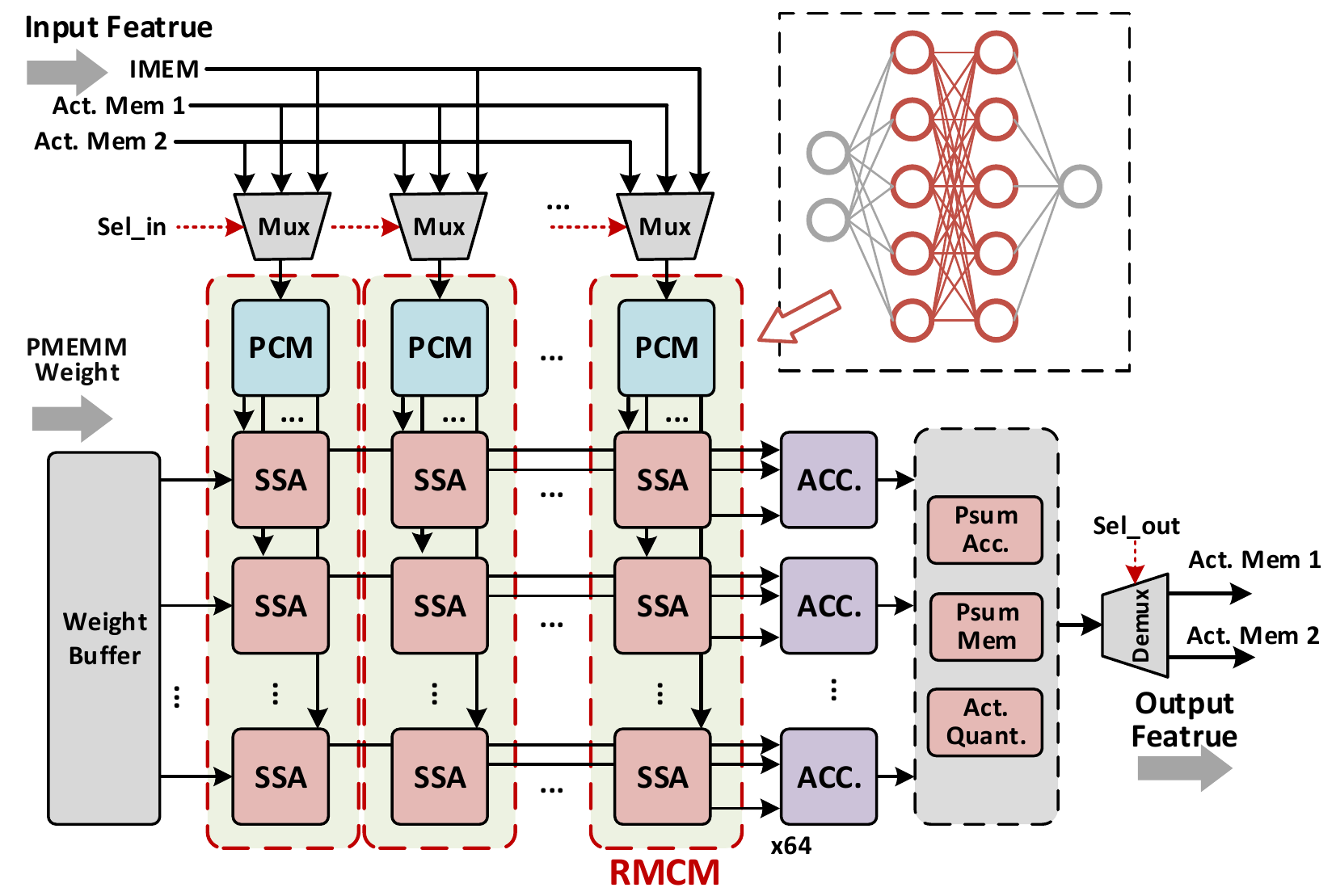}
    \caption{Computation of matrix-vector multiplication (MVM) using a multiple output network block (MONB).}
    \Description{}
    \label{fig:MOMB}
\end{figure}

\begin{figure*}[t]
    \centering
    \includegraphics[width=\textwidth]{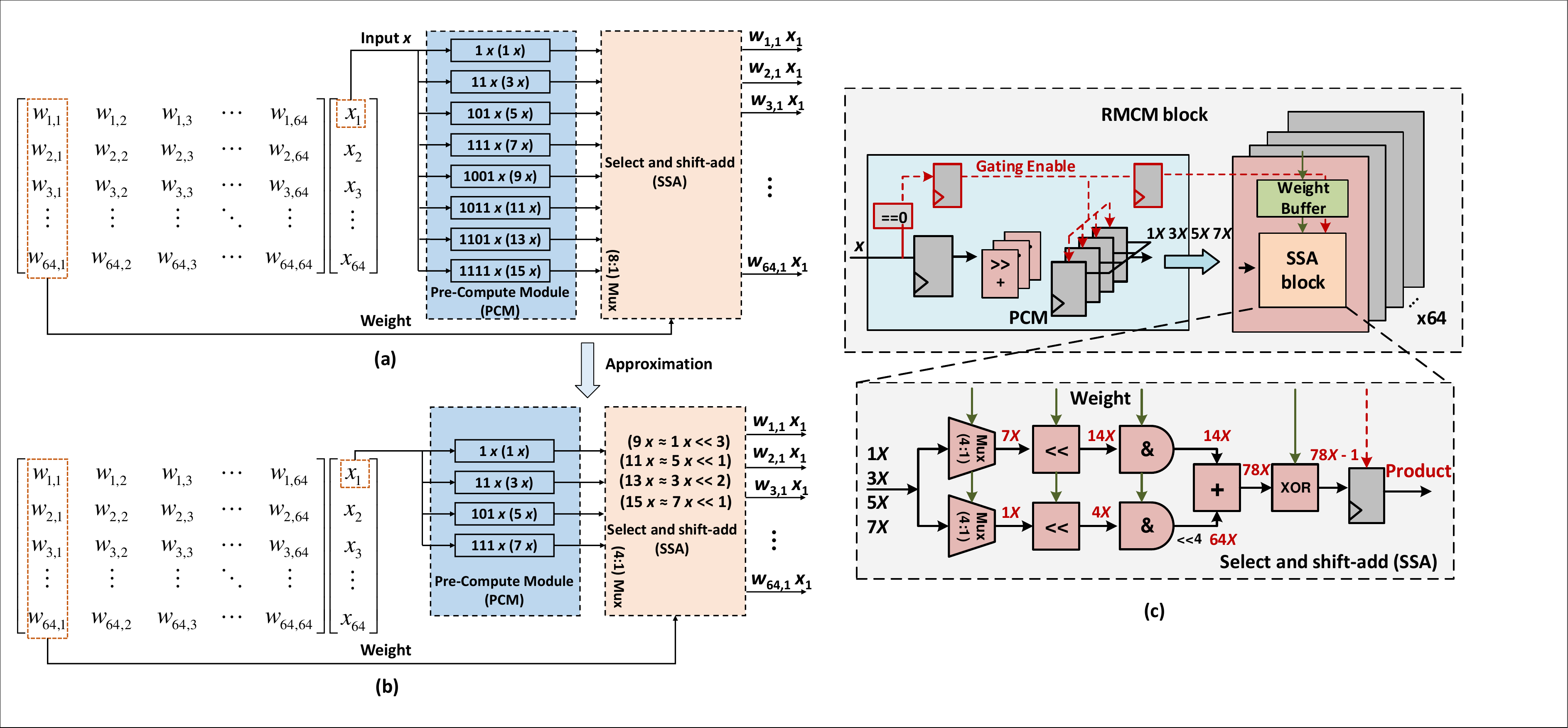}
    \caption{Working principle of (a) RMCM and (b) approximated RMCM. (c) Detailed implementation of the RMCM block. For approximated RMCM, the second half of common subexpressions are approximated using their nearest neighbors.}
    \label{fig:MCM_all}
\end{figure*}

Fig.~\ref{fig:computation_flow} illustrates the overall computation flow in an MLP engine. Instead of using a large matrix-vector multiplication (MVM) block for an entire layer, we partition it into several sub-MVM blocks with sizes of $64\times64$, making it more flexible for various NeRF-based rendering tasks with different matrix size (usually multiples of 64). For the implementation of a sub-MVM, taking the top-left block as an example, we transform it into 64 shift-add based reconfigurable multiple constant multiplication (RMCM) operations, where adder trees are further used to compress 64 partial results into the final resultant vector. The advantages of using RMCM are multi-fold. It is faster and more power-efficient than conventional MAC-based implementation since the common subexpressions are shared across different multipliers \cite{Park2004CSHM}. Moreover, we further adopt the idea of batch-computing to reduce off-chip memory access for weights. As illustrated in Fig.~\ref{fig:computation_flow}, after loaded into the sub-MVM block, the weight matrix stays stationary for a certain amount cycles and multiply with a set of input vectors for different sample points. Batch-computing reduces off-chip memory access for weight matrix at the cost of a small extra on-chip memory for intermediate output vectors. In our design, a batch size of 128 sample points are used, i.e., the weight matrix stays stationary in the MLP engine for the computation of sub-MVMs of 128 sample points.

\begin{figure}
    \centering
    \includegraphics[width=0.47\textwidth]{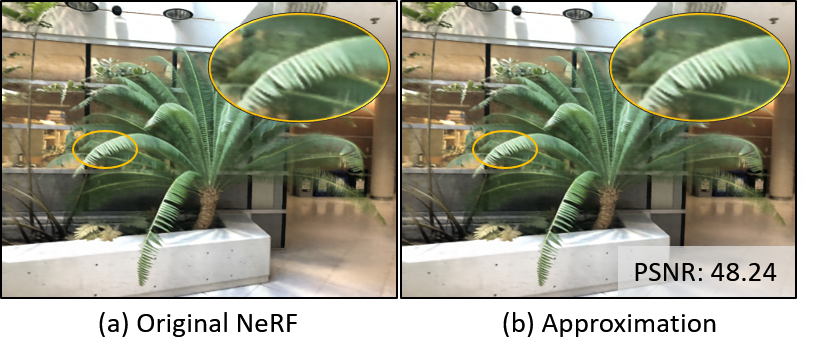}
    \caption{Two images rendered using (a) the original NeRF and (b) NeRF implemented using approximated RMCM.}
    \Description{Comparison of rendering results}
    \label{fig:rendering_appro}
\end{figure}

The computation of MVM in Fig.~\ref{fig:computation_flow} using a MONB block is illustrated in Fig.\ref{fig:MOMB}. In our design, a MONB block consists of 64 parallel RMCM blocks, where each RMCM block performs the RMCM operation shown in Fig.~\ref{fig:computation_flow} with 64 coefficients. The multiplication results of RMCM blocks at the same vertical positions, i.e., in the same row, are summed up by the corresponding adder tree. The summation results from the adder trees are further accumulated in the following block to generate the final MVM results. In particular, three buses load inputs come from the Input Memory, Activation Memory 1 and Activation Memory 2, where each data bus transfers 64 parallel activations. In each cycle, only one of the three data bus is activated by the multiplexers under the control of the Sel\_in signal. The selected data is fed to 64 pre-compute modules (PCMs), where each PCM generates 4 common subexpressions ($1x$, $3x$, $5x$ and $7x$) that shared by the following 64 select \& shift-add (SSA) blocks in the same column. Each SSA calculates the approximated product between input an activation and a weight by proper shift-add operations. Along the horizontal direction, the SSA in the same row computes 64 products of the same inner product operation, which are summed up by the adder tree to generate one inner product for the MVM. Once finished, the results of MVM are sent to the activation \& quantication (Act. Quant.) module for bias accumulation, activation operation and re-quantization operation. A demultiplexer (Demux) finally chooses the correct activation memory (Activation Memory 1 or Activation Memory 2) to save the output neurons.

Fig. \ref{fig:MCM_all}(a) illustrates the working principle of the RMCM for MVM computation, which is inspired by the work in \cite{Park2004CSHM}. The basic idea is to pre-compute a set of partial products, usually referred as common subexpressions, and share them across different multipliers. Since common subexpressions are fixed, they can be implemented using only shift-add operations. For example, $3x$ can be realized using only one shift-add operation as 

\begin{equation}
    \begin{split}
    3x = & 1x << 1 + 1x 
    \end{split}
\end{equation}
where "$<<$" represents a left-shift operation. In the select \& shift-add (SSA) block in Fig. \ref{fig:MCM_all}(a), multiplexers are used to select the needed common subexpressions according to the value of input weights. The selected common subexpressions are then shifted and summed up to generate the final multiplication results. Let us take a 9-bit weight $-78 (1\_0100\_1110)_2$ represented in the signed-magnitude form as an example. We first divide the weight into two parts, i.e., higher part $0100$ and lower part $1110$. Two multiplexers in the SSA block choose $1x$ and $7x$ for the higher part and lower part, respectively. Two shifters are then used to shift $1x$ and $7x$ to $0100x(4x)$ and $1110x(14x)$, respectively. Then we shift the result of higher part to the left by 4 bits to get $0100\_0000x (64x)$. An adder finally sums up the results of higher part and lower part to produce the final result $78x$.

Since neural networks are naturally fault-tolerant, we further adopt the idea of approximation to the RMCM block for further hardware complexity reduction. As illustrated in Fig. \ref{fig:MCM_all}(b), the second half of the pre-compute common subexpressions, i.e., $9X$, $11X$, $13X$ and $15X$, are omitted, and the values are approximated with their nearest neighbours. Since the number of common subexpressions is reduced from 8 to 4, the original (8:1) multiplexer is replaced by a (4:1), leading to hardware reduction for circuit implementation. In the meantime, the error introduced by this approximation is small (maximum error is 1/9 of the original multiplication result), which can be further compensated during the training process. Fig. \ref{fig:rendering_appro} shows two images rendered using (a) the original NeRF and (b) NeRF implemented using approximated RMCM. As we can see, there is no observable visual difference. The peak signal-to-noise ratio (PSNR) is as high as 48.24. By the introducing of approximation, the circuit area can be reduced by about 1/3.

Fig. \ref{fig:MCM_all}(c) illustrates the implementation details of the RMCM block. The pre-compute module generates the four common subexpressions, which are consumed by the following 64 SSA block. A comparator in PCM checks if the input is a zero or not, and gate the products (for zero input) for power-saving. A SSA block takes in the common subexpressions and generates the final multiplication result with proper selection and shift-add operations. 

\begin{figure}
    \centering
    \includegraphics[width=0.47\textwidth]{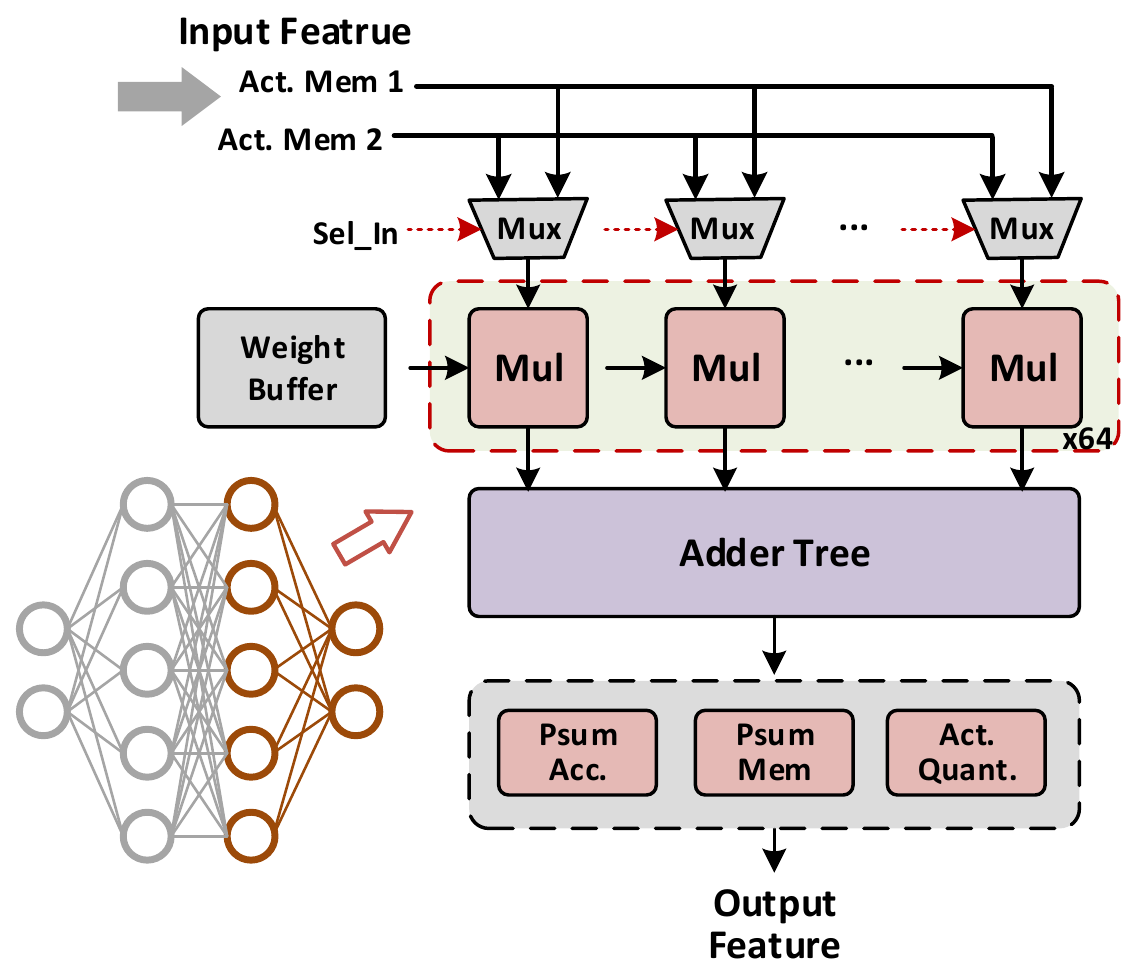}
    \caption{Detailed structure of the single output network block (SONB). It is used to compute the output layer of an MLP.}
    \Description{}
    \label{fig:SOMB}
\end{figure}

Compared with hidden layers, the computation of output layer is much simpler. In order to keep the same data from Activation Memory 1 and Activation Memory 2 to reduce the complexity of reading control, we implement 64 multiplication units in SONB as shown in Fig. \ref{fig:SOMB}. The data flow in SONB is similar to MONB except that the multiplications in SONB are implemented using general multipliers instead of RMCM blocks.

\subsection{Volume Rendering Unit}

In ICARUS, a VRU is implemented to render the final pixel colors. The main motivation of using a dedicated module instead of using CPU for volume rendering is to reduce the off-chip memory access. Since colors and densities of all sample points in a ray are compressed to the color of a pixel after volume rendering, data exchange between ICARUS and main memory is significantly reduced. Take the original NeRF as an example. To perform volume rendering outside ICARUS in a host CPU, we need to output the colors $\boldsymbol{c}=(r,g,b)$ and densities $\sigma$ of 128 sample points (fine rendering only) for each ray. While after volume rendering, the output data is the color $\boldsymbol{c}=(r,g,b)$ of one pixel, which is only $3/(128\times 4) \approx 0.586\%$ of that mentioned above.

The classic volume rendering equation \cite{volume-rendering,max_voluem_rendeirng} is given by
\begin{equation}\label{eq_volume_rendering}
\begin{aligned}
      C(r) = \sum^{N}_{i=1}T_i(1-\text{exp}(x_i))c_i, 
      \\ T_i =\text{exp}(\sum^{i-1}_{j=1}x_j), x_i = - \sigma_i\delta_i
\end{aligned}
\end{equation}
where $\delta_i = t_{i+1}-t_i$ is the distance between adjacent samples, $C(r)$ is the color of one ray we estimated by sample points, $\sigma$ is the volume density calculated by network and $c_i$ is the color of each sample point that goes through the activation function of sigmoid. We can rewrite the formula (\ref{eq_volume_rendering}) to
\begin{equation}\label{eq:volume_rendering2}
\begin{aligned}
    C(r) &= \sum^{N}_{i=1}(T_i-T_{i+1})c_i, 
   \\ T_{i+1} = \prod_{j=1}^{i-1}&\text{exp}(x_j)\cdot\text{exp}(x_i) = T_i\cdot \text{exp}(x_i)
 \end{aligned}
\end{equation}

\begin{figure}
    \centering
    \includegraphics[width=0.47\textwidth]{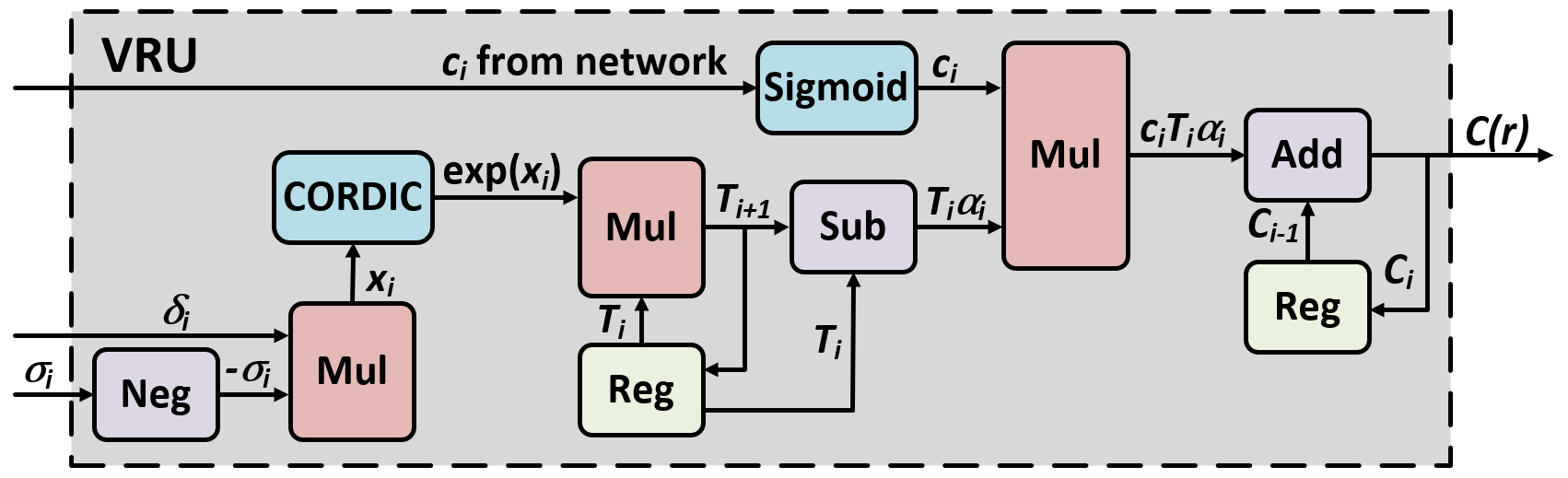}
    \caption{Detailed structure of the volume rendering unit (VRU).}
    \Description{}
    \label{fig:volume_rendering}
\end{figure}

Fig.\ref{fig:volume_rendering} illustrates the architecture of the VRU, which implements the rendering equation (\ref{eq:volume_rendering2}). In the VRU, we use a CORDIC module to calculate the exponential of $x_i$. The registered $T_i$ is used in two calculations: 1) the multiplication with exp($x_i$) to get $T_{i+1}$ and 2) the subtraction with $T_{i+1}$ to generate $T_i\cdot\alpha_i$, where $\alpha_i = 1 - \text{exp}(x_i)$. After finishing these two operations, $T_{i}$ is replaced by $T_{i+1}$ for the calculation in next accumulation. Once we have the coefficient of $c_i$, we multiply it with $c_i$ to get the weighted color $C_i$ of one sample point. The accumulations proceeds until we get the color of one pixel.

%% file: sections/implementation.tex
\section{Validation and Performance Evaluations}

\begin{figure}
    \centering
    \includegraphics[width=0.47\textwidth]{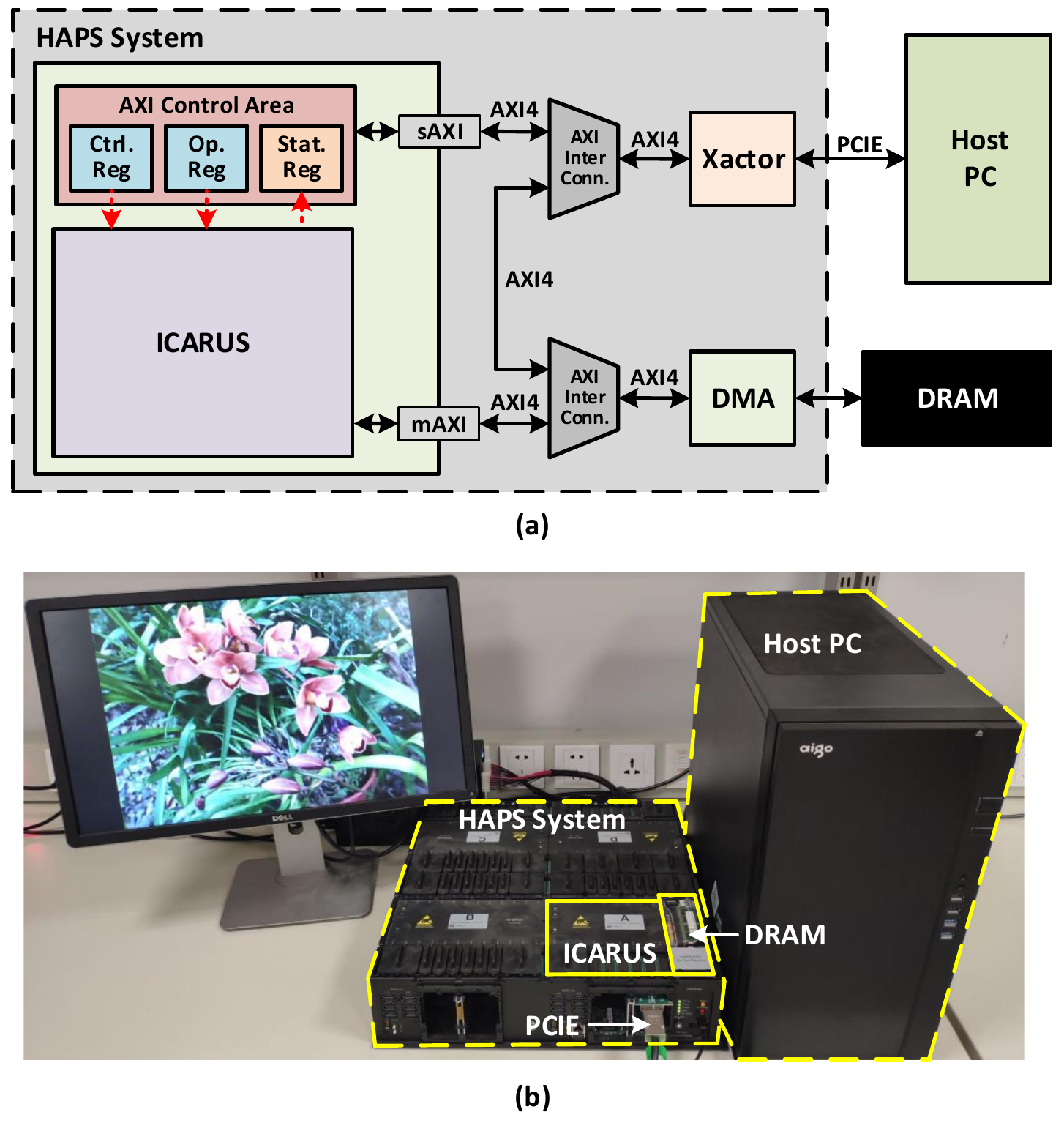}
    \caption{(a) Block diagram and (b) system setup of the FPGA-based prototype system. }
    \label{fig:FPGA_proto}
\end{figure}

To validate the proposed rendering system, we build a proof-of-concept prototype which implements 
a single core version 
ICARUS on Synopsys HAPS-80 S104, an FPGA-based prototype system.
Moreover, to evaluate the performance of ICARUS, we also map our design to 40nm CMOS technology.
The rendering performance of ICARUS, including the quality of rendered image as well as speed and energy efficiency, are compared with the GPU and TPU implementations.

\subsection{FPGA Prototype}

Fig. \ref{fig:FPGA_proto}(a) shows the block diagram of the FPGA prototype system, consisting of an ICARUS co-processor for NeRF rendering, a host PC for system control and a DRAM for data exchange between ICARUS and PC. The host PC connects to ICARUS through PCIe bus, where a Synopsys Xactor is used to decode the PCIe stream to AXI format. There are three addressable registers in the prototype system where two of them are write-only registers for saving processing instructions \& control signals and the other one is a read-only register for indicating the state of the system. The system setup is illustrated in Fig. \ref{fig:FPGA_proto}(b). After loading the network model, the host PC send inputs and instructions to the DRAM on HAPS. The ICARUS co-processor reads the inputs and processing instructions, and executes the NeRF pipeline. The rendered pixel colors are sent back to the host PC to generate the final image. The following evaluation results of NeRF and surface light field (SLF) rendering are performed using the above mentioned FPGA prototype.

\boldstartspace{Neural Radiance Field Rendering.} In our evaluation, for each pixel to render, we adopt a two-pass sampling strategy as in the original NeRF: first generate $64$ uniformly distributed samples within the visible range, calculate density distribution along the pixel ray and finally generate another $128$ samples that are more close to the surface of the object.

\begin{figure*}
    \centering
    \includegraphics[width=0.989\textwidth]{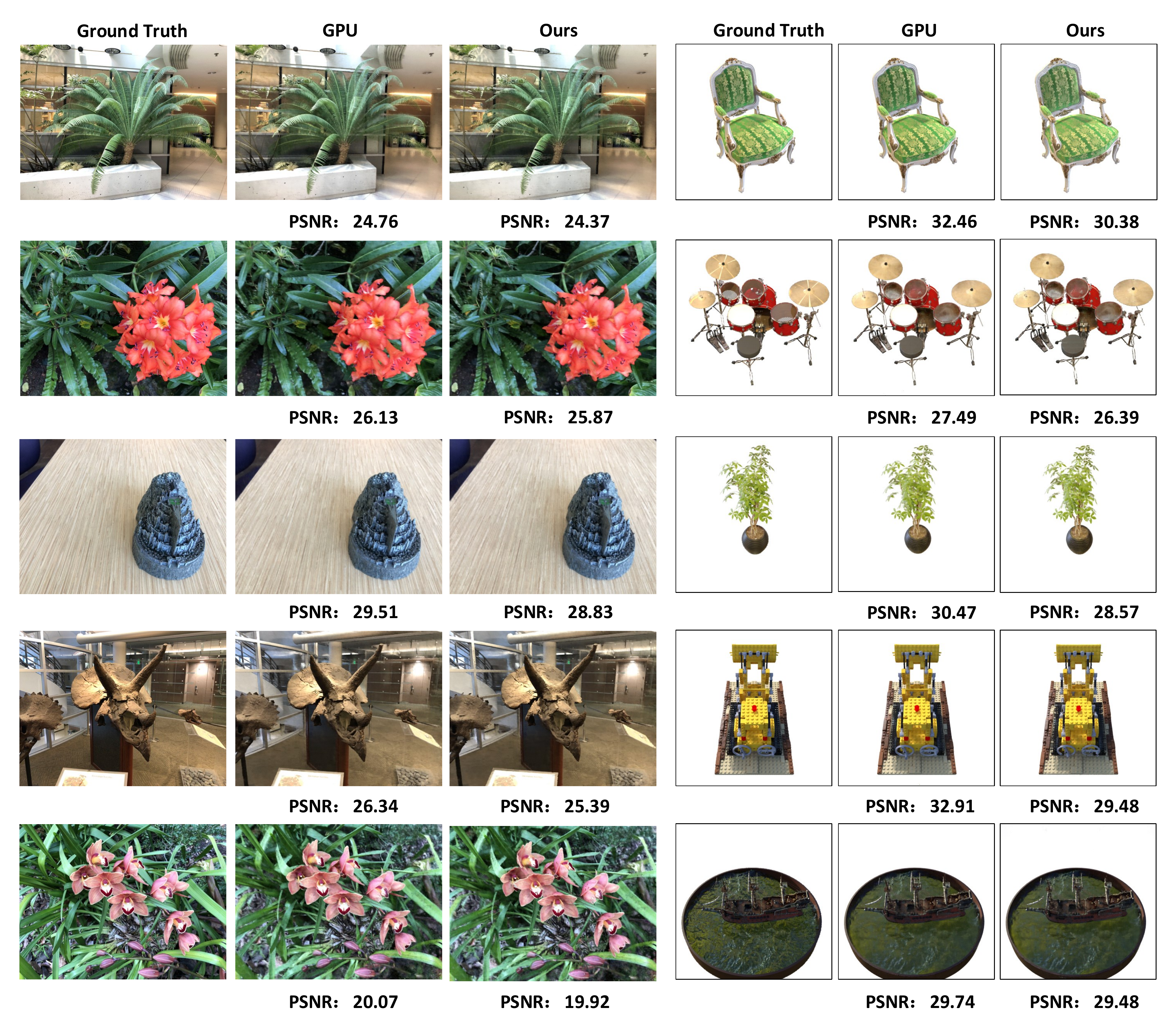}
    \caption{Comparison of NeRF rendering results using GPU and ICARUS. PSNRs are presented below each result.}
    \label{fig:nerf_resuls}
\end{figure*}

Fig. ~\ref{fig:nerf_resuls} compares the NeRF rendering results using GPU and ICARUS. Note that we also include the ground truth images for reference. 
We run comprehensive experiments using both synthetic and real scenes that are presented in previous works. As illustrated, there are practically no visual differences between our results and those generated by GPUs under all scenes. In the `ficus' and `fern' scene, we faithfully preserve fine leaf edges. Likewise, in the `ship' scene, we reproduce the reflectance of water and the thin mast of the ship. These experiments reveal the fact that fixed-point computation is appropriate for rendering tasks and validate the precision of our computing pipeline. They further demonstrate that our hardware is a sensible substitution of GPU in the domain of NeRF rendering.

In terms of quantitative evaluation, we use the commonly used PSNR to measure the quality of rendering results. As we can see, for most cases (especially the real scenes), the PSNRs of GPU results and ICARUS results are very close, where the gap is less than 1dB. But we also note that for certain cases in the synthetic scenes, for example the `chair', the PSNR gap between GPU results and ICARUS results can be as large as 2dB. This gap is caused by the post-training quantization we used in this work, where the weights are linearly quantized without retraining. But according to existing research on network quantization, this PSNR degradation can be alleviated or even complete addressed by some techniques such as quantization aware training ~\cite{quantization_aware}, which has been widely used in many neural network implementations.

\begin{figure*}
    \centering
    \includegraphics[width=0.95\textwidth]{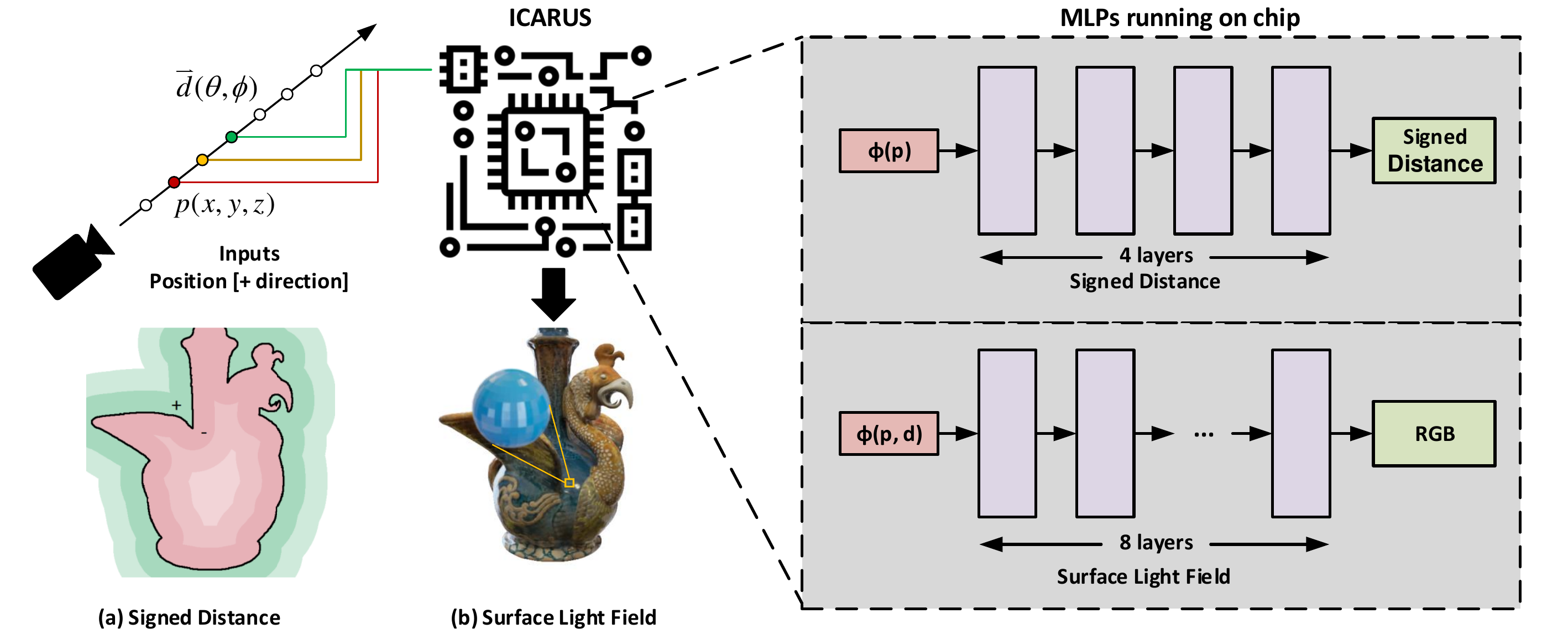}
    \caption{Computing architectures of (a) neural signed distance field and (b) neural surface light field.}
    \label{fig:net-arch}
\end{figure*}

\boldstartspace{Neural Surface Light Field Rendering.} A surface light field (SLF) ~\cite{wood2000surface} employs a collection of light rays that emit from the surface of an object towards all directions. Like NeRF, each light ray has an attached radiance value except that they are parametrized over the surface. In that perspective, an SLF serves as an object-level representation of the plenoptic function. Brute-force rendering using an SLF is straightforward: given a virtual camera in the free space, we directly trace view rays from its center of projection towards the surface, compute their interaction points with the scene, and the query SLF via techniques such as the unstructured lumigraph ~\cite{buehler2001unstructured}. For high quality rendering, a large number of images (at the same scale of NeRF) are required and hand-crafted blending schemes are essential to reduce visual artifacts.

Analogous to NeRF, deep surface light ~\cite{dslf} shows that one can also compactly encode an SLF into an MLP as shown in Fig. ~\ref{fig:net-arch}. Specifically, we leverage MLPs' abilities for providing continuous representations, delegating the interpolation duty to the neural network. Once the network is properly trained, we can treat it as an ordinary continuous function for querying novel radiance. Different from NeRF though, the input to the SLF network for querying the radiance is a six-dimensional vector: it consists of three elements for a spatial point on the object surface, and another three for the angular direction of the virtual ray. It is possible to further employ anisotropic Fourier features ~\cite{affm} to enhance view dependency.

We observe that an SLF exhibits strong variations in the spatial domain but low in the angular domain. Therefore we construct the transformation from two separate smoothing parameters, one for high spatial variance and the other for low angular variance. We use ICARUS to directly implement the corresponding transformations on the chip to reduce data transfers while achieving high performance. Unlike NeRF though, rendering an SLF requires explicit geometry such as a mesh. This indicates that ICARUS will have to interface with a full graphics pipeline, which is difficult to conduct in practice. Therefore, to bypass the GPUs by using the matrix-multiplication-based rendering on ICARUS, we adopt another neural network to handle geometry. Specifically, we use a signed distance function (SDF) network ~\cite{park2019deepsdf} that can implicitly encode any given mesh surfaces. In a nutshell, the SDF network takes in a spatial point in free space and produces the minimum distance from the point to the surface. Once converted, the SDF can be used to march any virtual ray to locate its intersection point with the surface through iterative evaluation using the SDF network.

\begin{figure}
    \centering
    \includegraphics[width=0.47\textwidth]{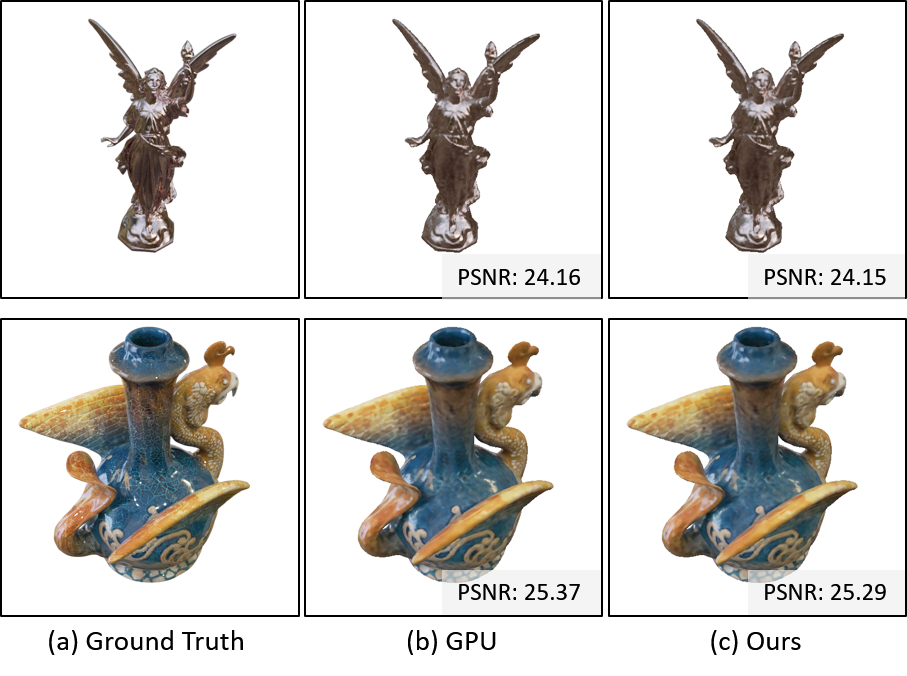}
    \caption{Comparison of GPU and ICARUS on the SLF rendering task.}
    \label{fig:slf_resuls}
\end{figure}

To train an MLP-based SLF (on the GPU), we employ the Fourier features to preserve maximum geometry details. Since an SDF is almost uniformly distributed and exhibits small variances across different axes, we use isotropic transformations using a single smoothing parameter. It is worth noting that our neural SLF rendering does not requires evaluating a large set of samples along each ray as in NeRF: once we obtain the intersection of any virtual ray with the surface, we only need to evaluate that single sample. Fig. \ref{fig:slf_resuls} shows a sample comparison between the GPU and the ICARUS for the neural SLF rendering task. Quality wise, both visually and quantitatively (in terms of PSNR), ICARUS achieves a similar performance to its GPU implementation.

\subsection{ASIC Evaluation}
To evaluate the speed and energy-efficiency of ICARUS, we map ICARUS system to ASIC platform based on 40nm CMOS technology. The design is implemented using Synopsys Design Compiler, and power consumption is estimated using Synopsys PrimePower based on real stimulus. To have a comprehensive comparison, we use different implementations of NeRF, including JaxNeRF, which is an optimized version of the original TF NeRF implementation~\cite{jaxnerf2020github}. Moreover, we also include implementation results of JaxNeRF using multiple devices.

\begin{table*}[t]
\renewcommand\arraystretch{1.5}
\centering
\caption{Comparison of NeRF implementation on GPU, TPU and ICARUS}
\label{asicResult}
\begin{tabular}{|c|c|cc|c|c|}
\hline
Implementation      & TF NeRF~\cite{jaxnerf2020github}          & \multicolumn{3}{c|}{JaxNeRF~\cite{jaxnerf2020github}}                       & Ours      \\ \hline
Platform            & NVIDIA V100 $\times$ 1         & \multicolumn{1}{c|}{NVIDIA V100 $\times$1}& \multicolumn{1}{c|}{NVIDIA V100 $\times$ 8} & TPUv2 $\times$ 128     & ICARUS        \\ \hline
Process                     & \multicolumn{3}{c|}{12 nm}           & 16 nm       & 40 nm       \\ \hline
Clock frequency                & \multicolumn{3}{c|}{1.245 GHz}        & 700MHz      & 400 MHz      \\ \hline
Image resolution     &  800 $\times$ 800       & \multicolumn{3}{c|}{800 $\times$ 800}       &  800 $\times$ 800            \\ \hline
Samples per ray     &  192       & \multicolumn{3}{c|}{192}       &  192              \\ \hline
Time for one frame          & 27.74 s         & \multicolumn{1}{c|}{20.77 s}& \multicolumn{1}{c|}{2.65 s}           & 0.35 s       & 45.75 s       \\ \hline
Silicon area                      & \multicolumn{2}{c|}{815 mm$^2$}& \multicolumn{1}{c|}{815 mm$^2$  $\times$ 8 }     & 611 mm$^2$ $\times$ 128 & 16.5 mm$^2$      \\ \hline
Power                   & \multicolumn{2}{c|}{300 W (TDP)}& \multicolumn{1}{c|}{300 W  $\times$ 8 (TDP)}     & 280 W $\times$ 128 (TDP) & 282.8 mW      \\ \hline
Energy-efficiency   &  -       & \multicolumn{1}{c|}{-}          & -      & - &   0.105 $\mu J/sample$     \\ \hline
\end{tabular}
\begin{tablenotes}
	    \item Energy-efficiency of other implementations is not given due to the lack of run-time power. 
	  \end{tablenotes}
\end{table*}

Table \ref{asicResult} presents the rendering performance comparison of NeRF on different platforms. To the best knowledge of the authors, ICARUS is the first dedicated hardware accelerator design for NeRF-based neural volume rendering tasks. As references, we compare the performance of ICARUS with GPU and TPU, for running the same NeRF algorithm with the same number of sample points per ray. 

As we can see, the rendering speed of ICARUS is lower than the TF NeRF and JaxNeRF implementation on a NVIDIA V100 GPU, but the silicon area and power consumption of ICARUS is much smaller. Moreover, the CMOS fabrication process we used is two generations older than that of GPU and TPUv2. We also note that the current implementation of ICARUS is running at a much lower clock frequency, i.e., 400MHz, which can be further improved through circuit-level optimization. 
It should be noted that as commercial products, GPU and TPU include many extra modules to support different applications. While on the other hand, ICARUS is a dedicated hardware architecture tailed for NeRF-type rendering. Therefore, it is hard to have a precise comparison for energy and (silicon) area efficiency since we cannot exclude unnecessary parts in the evaluation for GPU and TPU.

%% file: sections/discussion.tex
\section{Discussion and Future work}
Though ICARUS, as a specialized architecture, has advantages in energy and (silicon) area efficiency, there are a number of limitations in our current design. We discuss these limitations and possible solutions to be explored in this section.

\boldstartspace{Multi-core system. }
Although PLCores are power-efficient for NeRF rendering, a multi-core system is necessary to meet the requirement of real-time rendering. Due to the volumetric scene representation of NeRF, each sample point in a ray has to go through the pipeline while the processings are independent of one another. Accordingly, the PLCore architecture is designed to independently handle the NeRF processing pipeline of mapping positions \& directions to pixel colors. Therefore, NeRF rendering of an image using a multi-core ICARUS can be parallelized with a small overhead of control and off-chip memory exchange. But since all PLCores share the same data path that communicates with DRAM as shown in Fig. \ref{fig:overallSys}, the bandwidth of data path may become the bottleneck of the overall system as the number of core increased. Therefore, a memory architecture as well as a network-on-chip that not only provide sufficient data bandwidth for the multi-core system but also minimize the memory access should be carefully designed to support a multi-core ICARUS system.

\boldstartspace{Fixed-point representation and computation.} In our ICARUS design, we use the fixed-point number system, i.e., we convert a NeRF model pre-trained on GPUs to its fixed-point version to ICARUS via quantization. Such post-training quantization will inevitably introduce extra errors, leading to rendering quality degradation as shown in Fig. \ref{fig:nerf_resuls}. Therefore, it is essential to explore quantization-aware training techniques \cite{quantization_aware} that have previously shown great success on CNN models. Specifically, once we manage to integrate quantization into the training process, we will be able to search for the lowest bit-width of the weights of the MLP as we expect them to change under different complexity of the scene in both geometry and appearance. For example, our SLF rendering assumes known geometry and therefore the bit-width should be shorter than traditional NeRF. To that end, an extension to the ICARUS architecture that supports dynamic precision will greatly benefit computational efficiency and rendering quality in the final rendering.

\boldstartspace{Flexibility vs. efficiency.}
The neural rendering field has reignited interests on designing new data structures and scene representations. In fact we have observed significant and even accelerated efforts on improving the training and rendering sides of the neural radiance fields. Despite their heterogeneity in algorithm designs (e.g., spatial hashing vs. spatial embedding vs. Octree), it seems that two key components are still shared across all techniques, i.e., positional encoding (PE) and an MLP, shallow or deep. Recall that our ICARUS architecture does not aim for a fixed design. Rather, it hopes to support new configurations by recycling many key implementations such as the PEU and the MLP modules. In fact, the SLF render precisely demonstrates the possibility of combining auxiliary neural networks as well as the possibility of integrating directly into the GPU. Latest extensions to NeRF such as pure MLP based SIGNET \cite{SIGNET} for light field rendering can migrate onto ICARUS, with the PEU re-designed to support Gegenbauer Polynomials-based encoding. We also intend to explore implementing Instant-NGP on ICARUS which requires implementing hash-based encoding on board.

However, it is equally important to point out that tailored designs are not meant to be for general purposes: specially designed ray tracing accelerator will not be able to achieve comparable performance on rasterization tasks. This is the same case for our prototype ICARUS that requires striking a balance between efficiency and flexibility. For ICARUS, our hope is that its lightweight architecture will be more friendly for mobile devices, such as HMDs in AR/VR applications, smartphones, and 3D displays. In these scenarios, achieving photorealistic rendering on a light, small, low power, and most importantly, untethered HMDs are consumer pain points. In fact, it may be possible to couple ICARUS with any existing mobile GPU units where the background can be rendered via rasterization and the foreground via the neural radiance field.

\boldstartspace{Further extensions.}
Although ICARUS has largely focused on computer graphics and computer vision tasks, it may also be extended to broader fields such as medical imaging, electron microscopy and tomography, non-line-of-sight (NLOS) imaging, etc. What they share in common is the volume rendering model, core to the neural radiance field and its extensions. In fact volume rendering can be conducted in the Fourier space \cite{totsuka1993frequency} analogous to tomographic approaches in CT/MRI/EM/ET. For example, we are currently exploring neural CT reconstruction algorithms as well as how to export them onto ICARUS to support real-time slicing and volume rendering. Many NLOS imaging techniques have also followed the tomographic pipeline and require real-time scene recovery and rendering where neural approaches have emerged as a potential solution. But we understand that proper changes in different parts of ICARUS are necessary to support these aforementioned applications. Decisions need to be made upon the requirement of specific applications. We hope ICARUS will stimulate significant future developments for and beyond graphics and vision.

\begin{acks}
This work was supported by the Central Guided Local Science and Technology Foundation of China (YDZX202231000010001).
\end{acks}